\documentstyle[12pt]{article}

\begin{document}

%
% TITLE
%
\begin{titlepage}

\begin{flushright}
OU-HET 309 \\
hep-th/9812047 \\
December 1998
\end{flushright}
\bigskip
\bigskip

\begin{center}
{\Large \bf
Comments On Hamiltonian Formalism \\ 
Of $AdS/CFT$ Correspondence \\ 
}

\bigskip
Toshio Nakatsu and Naoto Yokoi\\
\bigskip
{\small \it
Department of Physics,\\
Graduate School of Science, Osaka University,\\
Toyonaka, Osaka 560, JAPAN
}
\end{center}
\bigskip
\bigskip
\begin{abstract}
As a toy model to search for Hamiltonian formalism of 
the $AdS/CFT$ correspondence,  
we examine a Hamiltonian formulation of 
the $AdS_2/CFT_1$ correspondence   
emphasizing unitary representation theory of the symmetry.  
In the course of a canonical quantization 
of the bulk scalars,  
a particular isomorphism between the unitary 
irreducible representations in the bulk and boundary 
theories is found. 
This isomorphism defines the correspondence of field operators. 
It states that field operators of the bulk theory are 
field operators of the boundary theory 
by taking their boundary values in a due way.  
The Euclidean continuation provides an operator formulation 
on the hyperbolic coordinates system. 
The associated Fock vacuum of the bulk theory 
is located at the boundary, 
thereby identified with the boundary CFT vacuum. 
The correspondence is interpreted as 
a simple mapping of the field operators 
acting on this unique vacuum. 
Generalization to higher dimensions 
is speculated. 
\end{abstract}

\end{titlepage}

%
%  TEXT
%

It has recently been conjectured \cite{Maldacena}
that supergravity (and string theory) 
on the universal cover of $AdS_{d+1}$ 
(often called $CAdS_{d+1}$) times a compact 
manifold is equivalent to 
a conformal field theory (CFT) living 
on the boundary of $CAdS_{d+1}$.
The agreement of the spectrum 
of supergravity fluctuations with the 
spectrum of operators in the conformal 
field theory provides some evidence 
for this conjecture 
\cite{H-O,F-F,Witten},\cite{Boer,M-F}. 
In particular 
correlation functions in the boundary 
CFT are derived 
\cite{Witten,GKP}
from the dependence of the supergravity 
action near the boundary.

In spite of these phenomenological evidence 
a rationale of this $AdS/CFT$ correspondence seems 
to be little revealed.   
Frameworks for the Hamiltonian formulation 
are proposed \cite{Witten},\cite{BDHM} 
to obtain an insight on it,  
but, presumably due to a complication of 
representation theory of the symmetry of 
$AdS_{d+1}$ for general $d$ and our 
little knowledge of higher dimensional CFT, 
still make it mysterious.

In this article, to obtain the clue,  
avoiding such difficulties, 
we investigate the Hamiltonian or operator formalism 
of the $AdS_2/CFT_1$ correspondence.  
Since the dimensionality itself seems 
\cite{Witten} irrelevant for the correspondence, 
one can expect that study of this toy model 
provides some insight on our understanding 
of the $AdS/CFT$ correspondence.

     First we examine the canonical quantization 
of massive scalar fields on $CAdS_2$ 
emphasizing the following two perspectives; i) 
unitary representation theory of symmetry group, 
and ii) normalizability of wave functions. 
It turns out that 
these two perspectives are not equivalent 
in the process of quantization 
although they could become consistent with each other.  
This is because 
at some region of $mass^2$ 
the normalizable wave functions constitute 
two unitary irreducible representations different from each 
other. Therefore classification of the scalar particles 
by the symmetry requires another means more than  
their mass. This phenomenon turns out to be 
a manifestation of the $AdS/CFT$ correspondence.  
As its explanation, 
an isomorphism between the unitary irreducible 
representations in the bulk and boundary theories 
is constructed in a specific manner. 
This isomorphism also shows that 
field operators of the bulk theory are  
field operators of the boundary theory 
by taking their boundary values in a due way. 
This picture of the $AdS/CFT$ correspondence is supplemented 
by an argument on the Euclidean theory. 

           The Euclidean continuation provides an operator 
formulation on the two-dimensional hyperbolic space. 
In particular it is defined on the hyperbolic coordinate system. 
The associated Fock vacuum is located at the boundary of the 
hyperbolic space. Therefore it is identified with the vacuum 
of the boundary theory. With this identification  
both field operators of the bulk and boundary theories  
can act simultaneously on the unique vacuum. 
The correspondence now becomes the aforementioned 
simple mapping of the field operators.

             Finally, we address two observations. 
One is related with the path-integral argument of the 
Euclidean theory based on the parabolic or elliptic 
coordinate system. The other is a speculation on 
the generalization to higher dimensions.

~

Note added : The $AdS_2/CFT_1$ correspondence  
is considered also in \cite{Strominger} 
from the different viewpoint and mapping of 
some bulk operators to the boundary theory, 
which are not equivalent to ours, 
is introduced there.

~

\subsection*{Preliminary on $AdS_2$}

     $1+1$ dimensional anti-de Sitter spacetime 
$AdS_2$ is topologically $S^1 \times {\bf R}$ and described by 
the metric 
\begin{eqnarray}
ds^2 
= \frac{-(dt)^2+(d\rho)^2}{\cos^2 \rho},
\label{cads2 metric}
\end{eqnarray}
The ranges of the coordinates $t$ and 
$\rho$ are $0 \leq t < 2\pi$ and $-\pi/2 < \rho < \pi/2$. 
The Minkowski time $t$ is periodic. 
The universal cover $CAdS_2$  
is topologically ${\bf R} \times {\bf R}$.
The range of $t$ becomes $-\infty < t < +\infty$.  
The isometry group is $SU(1,1)$
\footnote{
$SU(1,1)$ 
$=\left\{ \left( 
\begin{array}{cc} 
\alpha & \beta \\ 
\bar{\beta} & \bar{\alpha} 
\end{array}
\right),~~
\| \alpha \|^2-\| \beta \|^2=1 \right\}$.}. 
The group action may be conveniently described in  
the lightcone coordinates $(u,v)$.  
They are introduced by 
$u=e^{i(t+\rho)}$ and $v=e^{i(t-\rho)}$. 
$SU(1,1)$ acts on $(u,v)$ in the following manner  
\begin{eqnarray}
\left( \begin{array}{cc} 
   \alpha & \beta \\ 
   \bar{\beta} & \bar{\alpha} 
\end{array}\right) 
\circ 
(u,v)=
\left(\frac{\alpha u+\beta}{\bar{\beta}u+\bar{\alpha}},
\frac{\alpha v-\beta}{-\bar{\beta}v+\bar{\alpha}}
\right).
\label{SU(1,1) on ADS2}
\end{eqnarray}
One can easily see that this group action  
preserves the $AdS_2$ metric $dudv/(u+v)^2$ 
in the lightcone coordinates.
$AdS_2$ is a homogeneous space of $SU(1,1)$. 
We want to describe it explicitly for a later convenience. 
Let us introduce the following group elements 
\begin{eqnarray}
&&
k_t \equiv 
\left( \begin{array}{cc} 
   e^{it/2} & ~ \\ 
   ~        & e^{-it/2} 
\end{array}\right),~~~
g_x \equiv  
\left( \begin{array}{cc} 
   \cosh x/2 & i \sinh x/2 \\ 
   -i \sinh x/2 & \cosh x/2 
\end{array}\right), 
\nonumber \\
&&
a_{\sigma} \equiv 
\left( \begin{array}{cc} 
   \cosh \sigma/2 & \sinh \sigma/2 \\ 
   \sinh \sigma/2 & \cosh \sigma/2 
\end{array}\right). 
\label{KGN of SU(1,1)}
\end{eqnarray}
The lightcone coordinates 
are related with the coordinates $(t,\rho)$ 
through the group action,  
$(u,v)=k_t~g_{x(\rho)}\circ (1,1)$. 
Here $x(\rho)$ is determined by   
$\cosh x(\rho)= 1/\cos \rho$.  
Notice that the little group at $(u,v)=(1,1)$  
consists of $a_{\sigma}$. 
Therefore $AdS_2$ is the homogeneous space 
$SU(1,1)/{\bf R}$.

         As the bases of the Lie algebra $su(1,1)$
we take the ones which generate $a_{\sigma},g_x$ and $k_t$ 
\footnote{
$a_{\sigma}=e^{\sigma X},
g_x=e^{x Y},k_t=e^{tZ}$}
\begin{eqnarray}
X \equiv 
\left( \begin{array}{cc} 
   ~ & 1/2 \\ 
   1/2 & ~ 
\end{array}\right),~~~
Y \equiv 
\left( \begin{array}{cc} 
   ~ & i/2 \\ 
   -i/2 &  ~ 
\end{array}\right),~~~
Z \equiv 
\left( \begin{array}{cc} 
   i/2 & ~ \\ 
   ~   & -i/2 
\end{array}\right).~~~
\label{basis of su(1,1)}
\end{eqnarray}
They satisfy the algebra, $[Z,X]=Y,[Z,Y]=-X,[X,Y]=-Z$. 
The Killing vectors corresponding to 
these generators are denoted by 
$\bf x,y,z$. They turn out to have the forms 
\begin{eqnarray}
&&
\mbox{\bf x}
=\cos t \sin \rho \partial_t
+\sin t \cos \rho \partial_{\rho},~~~~
\mbox{\bf y}
=\sin t \sin \rho \partial_t 
-\cos t \cos \rho \partial_\rho,
\nonumber \\
&&
\mbox{\bf z}
=-\partial_t.
\label{real killing on ads2}
\end{eqnarray}
We also introduce the complexified bases, 
$E \equiv X-iY, F \equiv -X-iY$ and 
$H \equiv -2i Z$. These satisfy the algebra, 
$[H,E]=2E,[H,F]=-2F,[E,F]=-H$. 
Denote the corresponding complexified Killing 
vectors by $\bf e$ ,$\bf f$ and $\bf h$. 
They have the forms 
\begin{eqnarray}
&&
\mbox{\bf e}
= i e^{-it}(\cos \rho \partial_{\rho}
     -i \sin \rho \partial_t),~~~
\mbox{\bf f} 
= i e^{it}(\cos \rho \partial_{\rho}
     +i \sin \rho \partial_t),
\nonumber\\
&&
\mbox{\bf h}
= 2 i \partial_t. 
\label{complex killing on ads2} 
\end{eqnarray}
The Casimir element $C$ of $su(1,1)$ is 
given by $C=(X^2+Y^2-Z^2)/2$. Insertion of the 
Killing vectors into $2C$ 
leads to the scalar laplacian 
$\Box_{AdS_2}$
$=\cos^2 \rho 
(-\partial_t^2+\partial_{\rho}^2)$  
on $AdS_2$.

The boundary of $AdS_2$ consists of two 
circles located at $\rho=\pm \pi/2$. In the lightcone 
coordinates it is described by $u=-v$. 
$SU(1,1)$ acts as one-dimensional 
conformal symmetry on the boundary. 
Consider the boundary circle at $\rho=\pi/2$. 
The Minkowski time $t$ provides the coordinate. 
The conformal transformation,  
$t \rightarrow t^g $, can be  
read from (\ref{SU(1,1) on ADS2}) as 
\begin{eqnarray}
e^{i(t^g+\frac{\pi}{2})} 
=
\frac{\alpha e^{i(t+\frac{\pi}{2})}+\beta}
{\bar{\beta}e^{i(t+\frac{\pi}{2})}+\bar{\alpha}},
\label{SU(1,1) on boundary}
\end{eqnarray}
where $g=\left( 
\begin{array}{cc} 
\alpha & \beta \\ 
\bar{\beta} & \bar{\alpha} 
\end{array}
\right)$.

\subsection*{Massive scalar field on $CAdS_2$}

      Consider a free scalar field theory on 
$CAdS_2$. Let us parametrize its mass by 
$m^2=s(s-1)$. The field equation for this 
scalar field $\Phi(t,\rho)$ is simply given by 
\begin{eqnarray}
\{ -\Box_{AdS_2}+s(s-1) \} 
\Phi(t,\rho)=0 .
\label{field equation}
\end{eqnarray}
Formal solutions of this equation 
can be described using the Gegenbauer function
\begin{eqnarray}
\phi^{\lambda,\pm}_{\omega}
=e^{\pm i \omega t}
(\cos \rho)^{\lambda}
C_{\omega-\lambda}^{\lambda}(\sin \rho)
\label{formal solutions}
\end{eqnarray}
where the parameter $\lambda$ is equal 
to $s$ or $1-s$, 
and $\omega$ is an arbitrary non-negative real 
number.
$C^{\lambda}_{\alpha}$ denotes the Gegenbauer 
function defined by the hypergeometric function
\begin{eqnarray}
C^{\lambda}_{\alpha}(z)
=\frac{\Gamma(\alpha+2\lambda)}{\Gamma(\alpha+1)\Gamma(2\lambda)}
F(\alpha+2\lambda,-\alpha,\lambda+\frac{1}{2};\frac{1-z}{2}),
\end{eqnarray}
which satisfies 
\begin{eqnarray}
(1- z^2)\frac{d^2 w}{dz^2}-(2\lambda+1)\frac{dw}{dz}
+\alpha(\alpha+2\lambda)w=0.
\end{eqnarray}
Notice that the formal solution 
$\phi^{\lambda,\pm}_{\omega}$ has the equivalent expression 
by the Legendre function of the first kind, 
$\phi^{\lambda,\pm}_{\omega}$
$\sim$ 
$e^{\pm i \omega t}(\cos \rho)^{1/2} 
P^{\frac{1}{2}-\lambda}_{\omega-\frac{1}{2}}(\sin \rho)$.

         We start with the discussion on the normalizability 
of the formal solutions. For this purpose we introduce the field theoretical 
scalar product $(~,~)$  on the solutions of (\ref{field equation}) by 
\begin{eqnarray}
(\phi,\psi) 
\equiv 
-i \int_{-\pi/2}^{\pi/2}d\rho 
\{ \phi(t,\rho)\partial_t \bar{\psi}(t,\rho)
-\partial_t \phi(t,\rho) \bar{\psi}(t,\rho) \},
\label{scalar product}
\end{eqnarray}
where $\bar{\psi}$ is the complex conjugation of 
$\psi$. This scalar product is formally independent of 
$t$.

                     As one may easily observe it, 
the formal solutions $\phi^{\lambda,\pm}_{\omega}$ 
with general values of $\lambda$ and $\omega$
are not normalizable with this scalar product. 
If one requires the normalizability on the wave funtion, 
the allowed values of the parameters are restricted. 
In fact, the $\cos \rho$-factor in  $\phi^{\lambda,\pm}_{\omega}$ 
imposes the condition : 
$\lambda$ is real and greater than $-1/2$. 
At the same time the property of the Gegenbauer function 
requires the quantization of the energy $\omega$ : 
$\omega=n+\lambda$ where $n=0,1,2,\cdots$.
All the formal solutions $\phi^{\lambda,\pm}_{\omega}$ 
which satisfy these two conditions are normalizable. 
Since $\lambda$ is equal to $s$ or $1-s$, the above condition 
on $\lambda$ implies that $s$ must be real, which means 
$m^2 \geq -1/4$.

                     Normalizable solutions 
$\phi^{\lambda,\pm}_{n+\lambda}$ provides 
representations of $su(1,1)$. Denote the spaces spanned 
by these wave functions by ${\cal H}^{\lambda}_{\pm}\equiv$ 
$\bigoplus_{n \geq 0} \mbox{\bf C}\phi^{\lambda,\pm}_{n+\lambda}$.
The unitarity of the representation may be examined by looking at 
the group action on the orthonormal bases of the representation space.
They can be given by the following functions 
and their complex conjugations 
\begin{eqnarray}
\phi^{\lambda}_n(t,\rho)
\equiv 
c(\lambda)\sqrt{\frac{n!}{\Gamma(n+2\lambda)}}
e^{-i (n+\lambda)(t+\pi/2)}(\cos \rho)^{\lambda}
C^{\lambda}_n(\sin \rho),
\label{ortho wave function}
\end{eqnarray}
where $c(\lambda)\equiv \Gamma(\lambda)2^{\lambda-1}/\sqrt{\pi}$ 
is the normalization constant independent of $n$.
Due to the orthogonality  
\footnote
{$\int_{-1}^{1}dx(1-x^2)^{\lambda-1/2}C^{\lambda}_m(x)
C^{\lambda}_n(x)=
\frac{\pi \Gamma(n+2\lambda)}
{2^{2\lambda-1}n!(\lambda+n)\Gamma(\lambda)^2}
\delta_{m,n}$.}
of the Gegenbauer polynomials,  
$\phi^{\lambda}_m$ and $\bar{\phi}^{\lambda}_n$ satisfy 
the relations 
\begin{eqnarray}
(\phi^{\lambda}_m, \phi^{\lambda}_n) 
&=&\delta_{m,n},~~~
(\bar{\phi}^{\lambda}_m, \bar{\phi}^{\lambda}_n) 
= -\delta_{m,n}, 
\nonumber\\ 
(\phi^{\lambda}_m, \bar{\phi}^{\lambda}_n) 
&=& 0.
\label{ortho relation}
\end{eqnarray}
The orthonormal bases of 
${\cal H}^{\lambda}_{+}$ and 
${\cal H}^{\lambda}_{-}$ are  
respectively given by 
$\{\phi^{\lambda}_n\}_{n \geq 0}$ 
and 
$\{\bar{\phi}^{\lambda}_n \}_{n \geq 0}$.
The representations can be read from 
the action of the complex Killing vectors 
(\ref{complex killing on ads2}) on these orthonormal 
wave functions
\begin{eqnarray}
\mbox{\bf e}~\phi^{\lambda}_n
&=& \sqrt{(n+1)(n+2\lambda)}~\phi^{\lambda}_{n+1},~~~
\mbox{\bf f}~\phi^{\lambda}_n
= \sqrt{n(n-1+2\lambda)}~\phi^{\lambda}_{n-1},
\nonumber \\
\mbox{\bf h}~\phi^{\lambda}_n
&=& 2(n+\lambda)~\phi^{\lambda}_{n}. 
\nonumber \\ 
&& ~~~ 
\nonumber \\
\mbox{\bf e}~\bar{\phi}^{\lambda}_n
&=& -\sqrt{n(n-1+2\lambda)}~\bar{\phi}^{\lambda}_{n-1},~~~
\mbox{\bf f}~\bar{\phi}^{\lambda}_n
= -\sqrt{(n+1)(n+2\lambda)}~\bar{\phi}^{\lambda}_{n+1},
\nonumber \\
\mbox{\bf h}~\bar{\phi}^{\lambda}_n
&=& -2(n+\lambda)~\bar{\phi}^{\lambda}_{n}.
\label{unitary representation on ads2}
\end{eqnarray}
These explcit forms show that 
when $\lambda > 0$ 
the representations on ${\cal H}^{\lambda}_{\pm}$ are 
irreducible and unitary.
They can be thought as slight 
generalizations of the discrete series of the unitary irreducible 
representations of $SU(1,1)$ \cite{Sugiura}.
(Actually, if one restricts $\lambda$ to positive integers, 
they constitute the discrete series.)
On the other hand, when $-1/2< \lambda \leq 0$, 
the representations on ${\cal H}^{\lambda}_{\pm}$ 
are not unitary. 
This can be read from 
eqs.(\ref{unitary representation on ads2})  
related with $\phi^{\lambda}_0$ and $\bar{\phi}^{\lambda}_0$. 
Since $SU(1,1)$ is the symmetry group of $CAdS_2$
we require that stable scalar particles on it should belong 
to the unitary representations of $su(1,1)$.
This puts a further physical restriction 
on the allowed value of $\lambda$. It must satisfy $\lambda >0$.

~

      From the perspective of the above representation theory  
the allowed space of the normalizable solutions of 
(\ref{field equation}) can be summarized as follows 
\footnote{Due to the symmetry of $s$ in $m^2$ 
we can assume $s \geq 1/2$}:  
In the case of $s \geq 1$, 
it is given by 
${\cal H}^s\equiv$ ${\cal H}^s_+ \oplus {\cal H}^s_- $. 
Here ${\cal H}^s_{\pm}$ respectively 
consist of positive and negative energy modes. 
Therefore the scalar field $\Phi$ belongs 
to one unitary irreducible representation of $su(1,1)$.
In the case of $1/2 \leq s <1$, 
the allowed space of the normalizable 
solutions is ${\cal H}^{s}\oplus {\cal H}^{1-s}$. The scalar field 
$\Phi$ belongs to two different unitary irreducible representations .  
If one classifies the scalar particles on $CAdS_2$ by its symmetry 
these two unitary irreducible representations should be 
distinguished. But, as we have seen it, the consideration  
only on the field equation is insufficient to do this.

              This incompleteness of the discrimination of particles 
is a manifestation of the $AdS/CFT$ correspondence. 
To explain why it is, we first observe the boundary behaviors 
of the orthonormal wave functions 
$\phi^{\lambda}_n,\bar{\phi}^{\lambda}_n$ ($\lambda>0$). 
Of course they behave as $\sim (\cos \rho)^{\lambda}$ 
and approach to zero near the boundary of $CAdS_2$. 
But what we want to know is how they approach to zero.  
At the boundary $\rho=\pi/2$ it may be captured as follows  
\footnote{$C^{\lambda}_n(1)
=\frac{\Gamma(n+2\lambda)}{\Gamma(2\lambda)n!}$.} 
\begin{eqnarray}
\lim_{\rho \rightarrow \pi/2}
\frac{\Gamma(2\lambda)}{c(\lambda)}
(\cos \rho)^{-\lambda}
\phi^{\lambda}_n(t,\rho)
&=&
\beta^{\lambda}_n(t), 
\nonumber\\
\lim_{\rho \rightarrow \pi/2}
\frac{\Gamma(2\lambda)}{c(\lambda)}
(\cos \rho)^{-\lambda}
\bar{\phi}^{\lambda}_n(t,\rho)
&=&
\alpha^{\lambda}_n(t),
\label{mapping bulk to boundary}
\end{eqnarray}
where
\begin{eqnarray}
\beta^{\lambda}_n(t) 
\equiv 
\sqrt{\frac{\Gamma(n+2\lambda)}{n!}}
e^{-i(n+\lambda)(t+\frac{\pi}{2})},~~~~
\alpha^{\lambda}_n(t) 
\equiv 
\bar{\beta}^{\lambda}_n(t).
\label{ortho basis of boundary}
\end{eqnarray}
Due to the parity of the Gegenbauer function,
$C^{\lambda}_n(-x)=(-)^nC^{\lambda}_n(x)$, 
the description at the other boundary  
reduces to the above ones. 
In the next section we will see that  
$\{\beta^{\lambda}_n\}_{n \geq 0}$ and 
$\{\alpha^{\lambda}_n\}_{n \geq 0}$
constitute the orthonormal bases of the unitary 
irreducible representations of $su(1,1)$ 
living on the boundary. 
We assume this result for a while. 
Denote the spaces spanned by these boundary 
wave functions by 
${\cal W}^{\lambda}_+ \equiv 
\bigoplus_{n \geq 0} \mbox{\bf C} \beta^{\lambda}_n$
and 
${\cal W}^{\lambda}_- \equiv 
\bigoplus_{n \geq 0} \mbox{\bf C} \alpha^{\lambda}_n$.
The L.H.S. of eqs.(\ref{mapping bulk to boundary}) 
define the isomorphism 
between the unitary irreducible representations  
living on $CAdS_2$ and its boundary 
\begin{eqnarray}
{\cal H}^{\lambda}_{\pm} 
\stackrel{\sim}{\longrightarrow} 
{\cal W}^{\lambda}_{\pm}.
\label{isomorphism from bulk to boundary}
\end{eqnarray}
For the scalar field $\Phi^{\lambda}$ 
which belongs to ${\cal H}^{\lambda}$, 
the image $\lim_{\rho \rightarrow \pm\pi/2}$ 
$(\cos \rho)^{-\lambda}\Phi^{\lambda}(t,\rho)$ describes 
a wave function on the boundary 
which belongs to ${\cal W}^{\lambda}
\equiv {\cal W}^{\lambda}_+\oplus {\cal W}^{\lambda}_-$.

\subsection*{Boundary CFT revisited}

      Here we simply describe the unitary irreducible representations 
of $su(1,1)$ on the boundary of $CAdS_2$. 
Related conformal quantum mechanical models are investigated 
in \cite{CQM}, and recently in \cite{CDKKTP}.

       We begin with the case of $AdS_2$.
Let us consider the boundary circle located at $\rho=\pi/2$. 
Let $\varphi (t)$ be a tensor field of degree $h$ on this circle. 
It will transform covariantly under the conformal group action 
\begin{eqnarray}
\varphi (t) (dt)^h =
\varphi  (t^g)  (dt^g)^h  
~~~~~~   g~ \in~  SU(1,1). 
\end{eqnarray} 
The (complexified) generators 
$E,F$ and $H$ act on $\varphi$ 
as the following differential operators 
$\hat{\mbox{\bf e}},\hat{\mbox{\bf f}}$ 
and $\hat{\mbox{\bf h}}$
\begin{eqnarray}
&&
\hat{\mbox{\bf e}}=
e^{-i(t+\frac{\pi}{2})}
(h+'‰ \partial_t),~~~~~
\hat{\mbox{\bf f}}=
e^{i(t+\frac{\pi}{2})}
(-h+i \partial_t),
\nonumber \\
&&
\hat{\mbox{\bf h}}=2i \partial_t.
\label{complex killing on circle}
\end{eqnarray}
The corresponding Casimir element becomes 
$C=h(h-1)/2$. We can expand $\varphi$ by the modes
$e^{-in(t+\frac{\pi}{2})}(\equiv \varphi_n)$ 
$(n \in \mbox{\bf Z})$. 
The representation can be written as  
\begin{eqnarray}
&&
\hat{\mbox{\bf e}}~\varphi_n
~=~(n+h)~\varphi_{n+1},~~~~~
\hat{\mbox{\bf f}}~\varphi_n~=~
(n-h)~\varphi_{n-1}, 
\nonumber \\ 
&&
\hat{\mbox{\bf h}}~\varphi_n
~=~2n~\varphi_n.
\label{rep Vh}
\end{eqnarray}
The cases of 
$h$ being positive integers are 
physically reasonable.
In these cases there appear the invariant subspaces 
which provide the highest or lowest weight representations 
of the conformal group. 
Let $h=l \in \mbox{\bf Z}_{\geq 0}$.   
The subspaces,
$\bigoplus_{n\geq 0}
\mbox{\bf C}\varphi_{n+l}$ and 
$\bigoplus_{n\geq 0}\mbox{\bf C}\varphi_{-n-l}$ 
are invariant under the action (\ref{rep Vh}). 
Actually they are respectively  
${\cal W}^l_{+}$ and ${\cal W}^l_{-}$. 
Taking the bases $\beta^{l}_n$ and $\alpha^{l}_n$ 
the representations can be read as 
\begin{eqnarray}
\hat{\mbox{\bf e}}~\beta^{l}_n
&=& \sqrt{(n+1)(n+2l)}~\beta^{l}_{n+1},~~~
\hat{\mbox{\bf f}}~\beta^{l}_n
= \sqrt{n(n-1+2l)}~\beta^{l}_{n-1},
\nonumber \\
\hat{\mbox{\bf h}}~\beta^{l}_n
&=& 2(n+l)~\beta^{l}_{n}. 
\nonumber \\ 
&& ~~~ 
\nonumber \\
\hat{\mbox{\bf e}}~\alpha^{l}_n
&=& -\sqrt{n(n-1+2l)}~\alpha^{l}_{n-1},~~~
\hat{\mbox{\bf f}}~\alpha^{l}_n
= -\sqrt{(n+1)(n+2l)}~\alpha^{l}_{n+1},
\nonumber \\
\hat{\mbox{\bf h}}~\alpha^{l}_n
&=& -2(n+l)~\alpha^{l}_{n}.
\label{discrete series on boundary}
\end{eqnarray}
These coincide with the representation  
(\ref{unitary representation on ads2}) 
of the scalar field $\Phi^{\lambda=l}$.
In fact, introducing the hermitian norms 
such that $\beta^{l}_n$  and $\alpha^{l}_n$ 
are respectively orthonormal 
in ${\cal W}^l_+$ and ${\cal W}^l_-$, 
we obtain the unitary irreducible integrable 
representations of $su(1,1)$ which belong to 
the discrete series. 
Therefore the scalar field $\Phi^{\lambda=l}$ 
living on $AdS_2$  
can appear on the boundary as the tensor field 
$\varphi^l$ which belongs to ${\cal W}^l$ by 
\begin{eqnarray}
\varphi^l(t)
=\lim_{\rho \rightarrow \pi/2}
\frac{\Gamma(2l)}{c(l)}
(\cos \rho)^{-l}
\Phi^{l}(t,\rho). 
\end{eqnarray}

                 Now we discuss the case of $CAdS_2$. 
Consider the boundary {\bf R} located at $\rho=\pi/2$. 
Let $\varphi(t)$ be a tensor field of degree $s$ 
with the quasi-periodicity 
\begin{eqnarray}
\varphi(t+2\pi)=e^{-2\pi i s}\varphi(t).
\end{eqnarray}
(We can also regard $\varphi$ as a twisted tensor field 
on the boundary circle of $AdS_2$.) 
The generators $E,F$ and $H$
act on it as the differential operators 
$\hat{\mbox{\bf e}},\hat{\mbox{\bf f}}$ 
and $\hat{\mbox{\bf h}}$ replacing $h$ by $s$
in (\ref{complex killing on circle}). 
$\varphi$ and its complex conjugate  
$\bar{\varphi}$ are expanded respectively by 
$e^{-i(n+s)(t+\frac{\pi}{2})}$ and  
$e^{i(n+s)(t+\frac{\pi}{2})}$.
Taking the same route as in the previous case 
we can obtain 
${\cal W}^s_+$ and 
${\cal W}^s_-$ as the invariant subspaces.
They are spanned by $\beta^s_n$ and $\alpha^s_n$.   
On these invariant subspaces the differential operators 
act as 
\begin{eqnarray}
\hat{\mbox{\bf e}}~\beta^{s}_n
&=& \sqrt{(n+1)(n+2s)}~\beta^{s}_{n+1},~~~
\hat{\mbox{\bf f}}~\beta^{s}_n
= \sqrt{n(n-1+2s)}~\beta^{s}_{n-1},
\nonumber \\
\hat{\mbox{\bf h}}~\beta^{s}_n
&=& 2(n+s)~\beta^{s}_{n},
\nonumber \\
&& ~~~~ 
\nonumber \\
\hat{\mbox{\bf e}}~\alpha^{s}_n
&=& -\sqrt{n(n-1+2s)}~\alpha^{s}_{n-1},~~~
\hat{\mbox{\bf f}}~\alpha^{s}_n
= -\sqrt{(n+1)(n+2s)}~\alpha^{s}_{n+1},
\nonumber \\
\hat{\mbox{\bf h}}~\alpha^{s}_n
&=& -2(n+s)~\alpha^{s}_{n},
\end{eqnarray}
which coincide with the unitary irreducible 
non-integrable representations 
(\ref{unitary representation on ads2}). 
This means that 
the scalar field $\Phi^{s}$ on $CAdS_2$  
can appear on the boundary as 
the tensor field $\varphi^s$   
which belongs to ${\cal W}^s$ by 
\begin{eqnarray}
\varphi^s(t)
=\lim_{\rho \rightarrow \pi/2}
\frac{\Gamma(2s)}{c(s)}
(\cos \rho)^{-s}
\Phi^{s}(t,\rho). 
\end{eqnarray}

\subsection*{Canonical quantization of $\Phi^s$}

The canonical quantization of the scalar field $\Phi^s(t,\rho)$ 
will be performed following the standard recipe \cite{Birrell-Davies}. 
Using the orthonormal bases $\phi^s_n$ and $\bar{\phi}^s_n$,  
the field can be expanded as
\begin{eqnarray}
\Phi^s(t,\rho)=
\sum_{n \geq 0}a_n \phi^s_n(t,\rho)+
\sum_{n \geq 0}a^{\dagger}_n \bar{\phi}^s_n(t,\rho). 
\label{mode expansion of scalar}
\end{eqnarray}
The canonical quantization will be implemented by 
adopting the commutation relations
\begin{eqnarray}
[a_m,a_n^{\dagger}]~=~\delta_{m,n},~~~~
[a_m,a_n]~=~[a_m^{\dagger},a_n^{\dagger}]~=~0.
\label{CR}
\end{eqnarray}
The Fock vacuum and its dual are introduced by 
the conditions 
\begin{eqnarray}
~^{\forall} a_n ~|vac>~=~0,~~~~~~
<vac|~ ^{\forall} a^{\dagger}_n~=~0.
\label{Fock vacuum}
\end{eqnarray}
The generators of $su(1,1)$ are 
constructed from the Noether currents and 
realized 
on the Fock space 
${\cal F}^s\equiv
\bigoplus_{n_1,\cdots,n_k}
\mbox{\bf C} a_{n_1}^{\dagger} \cdots a_{n_k}^{\dagger}
|vac>$ by the following operators  
\begin{eqnarray}
&&
E ~=~
\sum_{n\geq 0} \sqrt{(n+1)(n+2s)}~a_{n+1}^{\dagger}a_n,~~~
F ~=~
\sum_{n\geq 0} \sqrt{n(n-1+2s)}~a_{n-1}^{\dagger}a_n,~~~
\nonumber \\
&&
H~=~\sum_{n\geq 0}   2(n+s)~a_{n}^{\dagger}a_n.~~~
\label{operators for su(1,1)}
\end{eqnarray}
We notice that the Fock vacuum is $su(1,1)$-invariant. 
This invariance might be thought as a trivial consequence 
of the canonical quantization procedure. But, its implication 
seems to be non-trivial. The isomorphism   
(\ref{isomorphism from bulk to boundary}) of the 
representations through the wave functions 
on the $CAdS_2$ and its boundary 
naturally leads to the identification of 
the Fock vacuum of the bulk theory 
and the vacuum of the boundary CFT. 
This means the $su(1,1)$-invariance of the Fock vacuum 
is nothing but the conformal invariance 
of the boundary CFT vacuum 
\footnote{We will investigate this 
observation again in the discussion 
on the Euclidean operator formalism.}.

     To check the consistency of our formulation 
of the $AdS_2/CFT_1$ correspondence 
let us examine the boundary behavior of 
the Wightman function ,
$<vac|$ $\Phi^s(t_1,\rho_1)$ 
$\Phi^s(t_2,\rho_2)$ $|vac>$.
Using the mapping (\ref{mapping bulk to boundary}) 
of the wave functions we can derive 
\begin{eqnarray}
&&
\lim_{\rho_1,\rho_2 \rightarrow \pi/2}
(\frac{\Gamma(2s)}{c(s)})^2 
(\cos \rho)^{-2s} 
<vac|\Phi^s(t_1,\rho_1)\Phi^s(t_2,\rho_2)|vac>
\nonumber \\
&&~~~~~~
=\sum_{n\geq 0}\beta^s_n(t_1)\alpha^s_n(t_2)
\nonumber \\
&&~~~~~~
=\Gamma(2s)
\frac{e^{is(t_1+t_2)}}{(e^{it_1}-e^{it_2})^{2s}}.
\label{two point function}
\end{eqnarray}
This is nothing but the two-point function 
$<\varphi^s(t_1)\varphi^s(t_2)>$ of 
the boundary conformal field $\varphi^s$. 
To present it in a familiar form, 
it is convenient to change the coordinate from $t$ to $x$ 
by $x=\tan t/2$. The conformal field $\varphi^s(t)$ transforms 
to $\hat{\varphi}^s(x)$ according to the relation 
$\varphi^s(t)(dt)^s=\hat{\varphi}^s(x)(dx)^s$. 
In particular it holds 
$<\hat{\varphi}^s(x_1) \hat{\varphi}^s(x_2)>$
$=<\varphi^s(t_1)\varphi^s(t_2)>$ 
$(\frac{dt_1}{dx_1})^s(\frac{dt_2}{dx_2})^s$.  
The last expression in (\ref{two point function}) 
leads to
\begin{eqnarray}
<\hat{\varphi}^s(x_1) \hat{\varphi}^s(x_2)>
=\Gamma(2s)
\frac{1}{[i(x_1-x_2)^{2s}]}.
\end{eqnarray}
This is the two-point function of the corresponding 
conformal quantum mechanics \cite{CQM}.

\subsection*{Operator formulation for Euclidean theory}

    We want to examine the operator formulation of the 
Euclidean theory by simply replacing the Minkowski time 
$t$ by $-i\tau$. 
The scalar field $\Phi^s$ now lives in the Euclidean $CAdS_2$.
The Euclideanization of $CAdS_2$ is two-dimensional hyperbolic 
space $H_2$, that is, the hyperbolic plane or the Poincare disk.
Notice that the Euclideanization of $AdS_2$ is not $H_2$, but 
the quotient space $H_2/{\bf Z}$, where the ${\bf Z}$-action 
can be derived from the periodicity of the time. 
To be explicit, we consider the hyperbolic plane, 
that is, the upper half-plane  
${\bf C}_+\equiv \{w=x+iy, y>0\}$ endowed with the 
Poincare metric, $\frac{(dx)^2+(dy)^2}{y^2}$. 
The coordinate $(\tau,\rho)$ of the Euclidean $CAdS_2$ are related 
with $(x,y)$ by $x=e^{-\tau}\sin \rho$ and 
$y=e^{-\tau}\cos \rho$. (With this transform the Poincare metric 
becomes $\frac{(d\tau)^2+(d\rho)^2}{\cos^2\rho }$.)
We assume that the mode expansion (\ref{mode expansion of scalar}) 
of $\Phi^s$ on $CAdS_2$ also provides the mode expansion on $H_2$ 
after the Euclidean continuation. Notice that the following behavior 
of $\phi^s_n$ and $\bar{\phi}^s_n$ as  
$\tau \rightarrow \pm \infty$
\begin{eqnarray}
&&
\lim_{\tau \rightarrow +\infty}\phi^s_n(\tau,\rho)=\infty,~~~
\lim_{\tau \rightarrow -\infty}\phi^s_n(\tau,\rho)=0,
\nonumber \\
&&
\lim_{\tau \rightarrow +\infty}\bar{\phi}^s_n(\tau,\rho)=0,~~~
\lim_{\tau \rightarrow -\infty}\bar{\phi}^s_n(\tau,\rho)=\infty. 
\end{eqnarray}
Since 
$\lim_{\tau \rightarrow +\infty}(x,y)=(0,0)$ 
and 
$\lim_{\tau \rightarrow -\infty}(x,y)=\infty$, 
we can see that 
the Euclidean positive (negative) energy modes 
$\phi_n^s$ $(\bar{\phi}_n^s)$ diverge at $(x,y)=(0,0)$ 
($(x,y)=\infty$) and converge to zero at 
$(x,y)=\infty$ ($(x,y)=(0,0)$). 
From the perspective of the Euclidean operator formalism 
this means that the Fock vacuum and its dual 
associated with the mode expansion are located at 
$\infty$ and $(0,0)$. They are on the boundary of $H_2$. 
This also implies the vacua of the bulk and boundary 
theories actually coincide.

           In the Euclidean theory the operators $E,F$ and 
$H$ given in (\ref{operators for su(1,1)}) 
will generate respectively the special conformal transformation, 
translation and dilation of one-dimensional conformal symmetry. 
Let us check this. The isometry group of the hyperbolic plane 
is $SL(2;{\bf R})$. The group action is given by 
$g \circ w= \frac{aw+b}{cw+d}$ where 
$g=$
$\left( 
\begin{array}{cc} 
a & b \\ 
c & d 
\end{array}
\right)$
$ \in SL(2;{\bf R})$.  Fixing the isomorphism 
$SL(2:{\bf R})$
$\stackrel{\sim}{\rightarrow}$
$SU(1,1)$ 
by $g \rightarrow$ 
$\gamma g \gamma^{-1}$,  
where  $\gamma=\frac{1}{1+i}\left( 
\begin{array}{cc} 
i & 1 \\ 
-i & 1 
\end{array}
\right)$, 
let us take 
$\gamma^{-1}X\gamma$, $\gamma^{-1}Y\gamma$ and 
$\gamma^{-1}Z\gamma$ as the bases of $sl(2:{\bf R})$.
Denote the corresponding Killing vectors 
on the hyperbolic plane by 
${\bf x}_{H_2},{\bf y}_{H_2}$ and ${\bf z}_{H_2}$. 
They have the forms 
\begin{eqnarray}
&&{\bf x}_{H_2}=-\partial_{\tau},~~~
{\bf y}_{H_2}= -\cosh \tau \cos \rho \partial_{\rho}
+\sinh \tau \sin \rho \partial_{\rho},
\nonumber \\  
&&{\bf z}_{H_2}
=-\sinh \tau \cos \rho \partial_{\rho}
+\cosh \tau \sin \rho \partial_{\rho}. 
\end{eqnarray}
After the Euclidean continuation 
the complex Killing vectors on $CAdS_2$ 
can be written in terms of
${\bf x}_{H_2},{\bf y}_{H_2}$ and ${\bf z}_{H_2}$ as 
\begin{eqnarray}
{\bf e}=-i({\bf y}_{H_2}-{\bf z}_{H_2}),~~~
{\bf y}=-i({\bf y}_{H_2}+{\bf z}_{H_2}),~~~ 
{\bf h}= 2{\bf x}_{H_2}.
\end{eqnarray}
Therefore the operators $E,F$ and $H$ represent 
$\gamma^{-1}(Y-Z)\gamma$, $\gamma^{-1}(Y+Z)\gamma$ 
and $2\gamma^{-1}X\gamma$ of $sl(2:{\bf R})$, 
which respectively generate the special conformal 
transform, translation and dilation on the boundary 
of the hyperbolic plane.

~

      These identifications of the vacua 
besides the field operators  
suggest that the Fock space of the bulk theory itself 
has an interpretation in terms of the boundary physics. 
Notice that the one particle states of $\Phi^s$ are
nothing but ${\cal W}^s_+$. It can be thought as the Hilbert space of 
one-body conformal quantum mechanics. The identification 
of the Fock space ${\cal F}^s$ with the infinite sum 
of the symmetric product of ${\cal W}^s_+$, namely  
${\cal F}^s=$
$\bigoplus_{l \geq 0}$
$sym ( \underbrace{{\cal W}^s_+ 
\times \cdots \times {\cal W}^s_+}_{l} )$, 
implies that the boundary theory is $\infty$-body conformal 
quantum mechanics  
rather than one-body.

\subsection*{Outlook}

Two observations are addressed.

\subsubsection*{(i) Comparison with path-integral formulation}

Our operator formulation of Euclidean theory depends on 
the coordinate $(\tau,\rho)$ of the hyperbolic space $H_2$. 
This coordinate system is hyperbolic in the following sense : It can 
be determined by the $SL(2:{\bf R})$ parametrization,  
$w=\gamma^{-1}a_{\tau}g_{x(\rho)}\gamma \circ i$,  of 
the upper half-plane 
and there $a_{\tau}$ is a hyperbolic element of $SU(1,1)$. 
While this, the path-integral arguments \cite{Witten,GKP} 
are based on the coordinates $(x,y)$ of the upper half-plane 
or the radial coordinates $(\theta,r)$ of the Poincare 
disk. These coordinate systems are respectively 
parabolic and elliptic in the sense that  they are realized 
by the $SL(2:{\bf R})$ parametrizations, 
$w=$ 
$\left( 
\begin{array}{cc} 
1 & x \\ 
0 & 1 
\end{array}
\right)$ 
$\left( 
\begin{array}{cc} 
y^{1/2} & 0 \\ 
0 & y^{-1/2} 
\end{array}
\right)\circ i$ 
and 
$w=\gamma^{-1}k_{\theta}a_{\sigma(r)}\gamma \circ i$
\footnote{$\sigma(r)=\ln \frac{1+r}{1-r}.$}. 
In the $CAdS_2$ case 
the path intgral argument of the $AdS/CFT$ correspondence 
may be summarized as follows : 
Take the hyperbolic plane 
with the parabolic coordinates $(x,y)$ for simplicity . 
Consider a tensor field of degree $s$ on the boundary. 
Let us denote it by $\varphi_0^s(x)$. 
It transforms covariantly under the 
$SL(2:{\bf R})$ 
\begin{eqnarray}
\varphi_0^{s}(x)(dx)^s=
\varphi_0^{s}(g \circ x)(d (g\circ x))^s~~~~~
g \in SL(2:{\bf R}).
\end{eqnarray} 
This defines a representation of $SL(2:{\bf R})$ 
on the space of $\varphi_0^s$. 
Denote it $V^s$. There exists a natural pairing 
between $\varphi_0^s$ and $\varphi_0^{1-s}$ 
\begin{eqnarray}
<\varphi_0^{1-s},\varphi_0^{s}>\equiv 
\int_{-\infty}^{+\infty}dx~
\varphi^{1-s}_0(x)\varphi^s_0(x),
\end{eqnarray}
which is interpreted \cite{Witten} as the interaction 
between the bulk and boundary theories.
Consider the generalized Poisson integral 
\cite{Helgason}, 
\begin{eqnarray}
\int_{-\infty}^{+\infty}dx' 
\left(\frac{y}{y^2+(x-x')^2}\right)^s 
\varphi^{1-s}_0(x'), 
\end{eqnarray}
which can be rewritten in terms of the representation 
theory as 
\begin{eqnarray}
<\varphi_0^{1-s},
V^s \left( 
\left( 
\begin{array}{cc} 
1 & x \\ 
0 & 1 
\end{array}
\right)
\left( 
\begin{array}{cc} 
y^{1/2} & 0 \\ 
0 & y^{-1/2} 
\end{array}
\right) 
\right)
\varphi_0^{s}> ,
\end{eqnarray}
where
$\varphi_0^{s}(u)\equiv 
\frac{1}{(1+u^2)^s}$. 
This provides a formal solution of the bulk scalar 
field equation with $m^2=s(s-1)$.  
It is divergent on the boundary. 
By the semi-classical argument using this 
solution the correct two-point function 
of the boundary theory is derived \cite{Witten,GKP} 
from the bulk theory action.  
Those problems of the unitarity of representations 
and the normalizability of wave functions, 
which play the crucial role in the operator 
formulation on the hyperbolic coordinate system, 
do not appear, 
at least, superficially. 
The physical picture 
of the correspondence becomes completely different 
depending on the coordinate system which one takes.
It is interesting to construct the Bogoliubov 
transforms between the vacua constructed on  
the coordinate systems of different types.

\subsubsection*{(ii) Generalization to higher dimensions}

As a simple application, 
we speculate on the case of three dimensions. 
$CAdS_3$ is topologically ${\bf R}\times {\bf R}^2$ and 
endowed with the $AdS_3$ metric 
$\frac{-(dt)^2+(d\rho)^2+ \sin^2\rho (d\theta)^2}{\cos^2 \rho}$.
The ranges of the coordinates are 
$-\infty < t <+\infty,~ 0 \leq \rho < \pi/2$ and 
$0 \leq \theta < 2\pi$. 
The identification, $t\sim t+2\pi$, 
yields $AdS_3$. 
The Euclideanization of $CAdS_3$ is 
the three-dimensional hyperbolic 
space $H_3$. Consider the three-dimensional upper half-space, 
$\{(x,y,z)~~z>0\}$ endowed with the hyperbolic metric 
$\frac{(dx)^2+(dy)^2+(dz)^2}{z^2}$. With the change of coordinates, 
$x=e^{-\tau}\sin\rho\cos\theta,~y=e^{-\tau}\sin\rho\sin\theta$ and 
$z=e^{-\tau}\cos\rho$, the metric acquires the form, 
$\frac{(d\tau)^2+(d\rho)^2+\sin\rho^2(d\theta)^2}{\cos^2\rho}$. 
So, it is actually the Euclideanization of $CAdS_3$.

In the hyperbolic coordinate system $(\tau,\rho,\theta)$, 
the operator formulation of the bulk scalars tells 
the Fock vacuum of the bulk theory 
is located at the boundary of $H_3$. 
The $so(2,2)$-invariance of the Fock vacuum of the bulk
theory automatically implies the $sl(2:{\bf C})$-invariance 
of the boundary $2d$ CFT vacuum  
by the identification of these two. 
Field operators of the bulk and boundary theories can  
act on this unique vacuum. We can expect that 
the $AdS/CFT$ correspondence is realized along the same 
line as in the case of $CAdS_2$. Namely, 
the bulk theory field operators become 
the boundary theory field operators  
by taking their boundary values   
in such a way to define  
the isomorphism between the unitary irreducible 
representations living on the bulk and the boundary.

%\newpage

~

~

\subsection*{Acknowledgement}
We thank to K. Ohta for his collaboration at the early 
stage of this work. T.N. also thanks to N. Ishibashi 
for his useful comment.

\end{document}